\documentclass[twocolumn,aps,prl,preprintnumbers,bibnotes10pt,superscriptaddress]{revtex4}

\usepackage{graphicx}
\usepackage{dcolumn}
\usepackage{epsfig}
\usepackage{bm}
\usepackage{array}
\usepackage{float}
\usepackage{amsmath}
\usepackage{lipsum}
\usepackage{color}
\usepackage{bbm}
\usepackage{ulem}
\usepackage{braket}

\definecolor{MainRed}{RGB}{69, 187, 52}


\begin{document}
\title{Collisions of self-bound quantum droplets}
\author{Giovanni Ferioli}
\author{Giulia Semeghini}
\affiliation{LENS and Dipartimento di Fisica e Astronomia, Universit\'a di Firenze, 50019 Sesto Fiorentino, Italy} 
\affiliation{CNR Istituto Nazionale Ottica, 50019 Sesto Fiorentino, Italy}
\author{Leonardo Masi}
\affiliation{LENS and Dipartimento di Fisica e Astronomia, Universit\'a di Firenze, 50019 Sesto Fiorentino, Italy} 
\affiliation{CNR Istituto Nazionale Ottica, 50019 Sesto Fiorentino, Italy}
\author{Giovanni Giusti}
\affiliation{LENS and Dipartimento di Fisica e Astronomia, Universit\'a di Firenze, 50019 Sesto Fiorentino, Italy} 
\author{Giovanni Modugno}
\author{Massimo Inguscio}
\affiliation{LENS and Dipartimento di Fisica e Astronomia, Universit\'a di Firenze, 50019 Sesto Fiorentino, Italy} 
\affiliation{CNR Istituto Nazionale Ottica, 50019 Sesto Fiorentino, Italy}
\author{Albert Gallem\'i}
\author{Alessio Recati}
\affiliation{INO-CNR BEC Center and Dipartimento di Fisica, Universit\`a di Trento, 38123 Povo, Italy}
\affiliation{Trento Institute for Fundamental Physics and Applications, INFN, 38123, Trento, Italy}
\author{Marco Fattori}
\affiliation{LENS and Dipartimento di Fisica e Astronomia, Universit\'a di Firenze, 50019 Sesto Fiorentino, Italy} 
\affiliation{CNR Istituto Nazionale Ottica, 50019 Sesto Fiorentino, Italy}

\date{\today}

\begin{abstract}
We report on the study of binary collisions between quantum droplets formed by an attractive mixture of ultracold atoms. We distinguish two main outcomes of the collision, i.e. merging and separation, depending on the velocity of the colliding pair. The critical velocity $v_c$ that discriminates between the two cases displays a different dependence on the atom number $N$ for small and large droplets. By comparing our experimental results with numerical simulations, we show that the non-monotonic behavior of $v_c(N)$ is due to the crossover from a compressible to an incompressible regime, where the collisional dynamics is governed by different energy scales, i.e. the droplet binding energy and the surface tension. These results also provide the first evidence of the liquid-like nature of quantum droplets in the large $N$ limit, where their behavior closely resembles that of classical liquid droplets.
\end{abstract}

\maketitle

Collisions between self-bound objects have been studied in the most diverse physical systems. The best known example is that of classical liquids. When they approach each other with a certain relative velocity, liquid drops can either merge in a single droplet (coalescence) or separate into two or more drops after collision, depending on whether or not the surface tension is sufficient to counteract the kinetic energy of the colliding pair \cite{ashgriz_poo_1990, qian_law_1997, pan_2005}. Analogous studies have been carried out in the context of atomic nuclei to understand the dynamics of nuclear reactions and fission \cite{andreyev_2018, bulgac_2016, magierski_2017, bulgac_2017}. In the latter, for example, the transition from a single compound nucleus to the formation of two separate nuclei is governed by the interplay of collective macroscopic effects, described by the so-called liquid-drop model, and single-particle microscopic effects, related to shell corrections and pairing. Quantum effects in the collision of liquid droplets have been observed in the coalescence of helium clusters, where the merging dynamics occurs on a faster timescale with respect to classical fluids, due to the vanishing viscosity in their superfluid bulk \cite{vicente_2000, maris_2003, ishiguro_2004}. 
In all of these cases the study of binary collisions has proved to be a powerful tool to probe the dynamical properties of self-bound systems.

In this Letter, we consider a new entry in this class of self-bound objects, i.e. quantum droplets formed by a mixture of ultracold atoms. They consist in dilute samples of bosonic atoms with attractive interactions, stabilized against collapse by the repulsive effect of quantum fluctuations \cite{lhy_1957, petrov_2015, arlt_2018}. This novel quantum phase was first predicted few years ago \cite{petrov_2015} and it has drawn increasing attention in the community since then. Besides being a macroscopic manifestation of quantum fluctuations, quantum droplets are predicted to display a number of interesting features. Despite being extremely dilute, for large number of atoms they enter a liquid-like incompressible phase, highlighted by a uniform bulk density. Another exotic property is related to their excitation spectrum in the small atom number regime, where the discrete collective excitations are much higher in energy than the particle emission threshold \cite{petrov_2015}, which gives rise to a self-evaporation mechanism that continuously cools the droplet, in close analogy with the decay of giant resonances in nuclei (see, e.g., \cite{chomaz_1987} and references therein).

The first experimental observations of quantum droplets in atomic mixtures have been reported recently \cite{cabrera_2018, cheiney_2018, semeghini_2018}. Besides proving the existence of a self-bound phase in a mixture of $^{39}$K atoms, they also presented a first characterization of its phase diagram and equilibrium properties. In order to further look into its peculiar nature, in this work we study collisions between two quantum droplets. Analogous experimental studies in the context of ultracold atoms have been performed on bright solitons \cite{khaykovich_2002, strecker_2002} and dipolar quantum droplets \cite{schmitt_2016, igor_2016, chomaz_2016}. While in the latter case the result of collisions is mainly determined by dipolar interactions and the droplets repel each other \cite{igor_coll_2016}, collisions of bright solitons in one-dimensional waveguides directly probe their intrinsic properties, being influenced by their relative phase or by deviations from one dimensionality \cite{nguyen_2014, khaykovic_2006}. Theoretical studies about collisions of self-bound atomic clouds have been carried out in \cite{astrakharchik_2018} and \cite{malomed_2D} for one and two-dimensional mixture droplets and in \cite{adhikari_2017} for analogous self-bound quantum balls stabilized by repulsive three-body interactions. In this Letter we show how the study of collisions between Bose-Bose mixture droplets is a powerful tool to verify the existence of a liquid regime at large atom numbers and to gain information about the energy scales of the system.

\begin{figure}[t] 
\begin{center}
\includegraphics{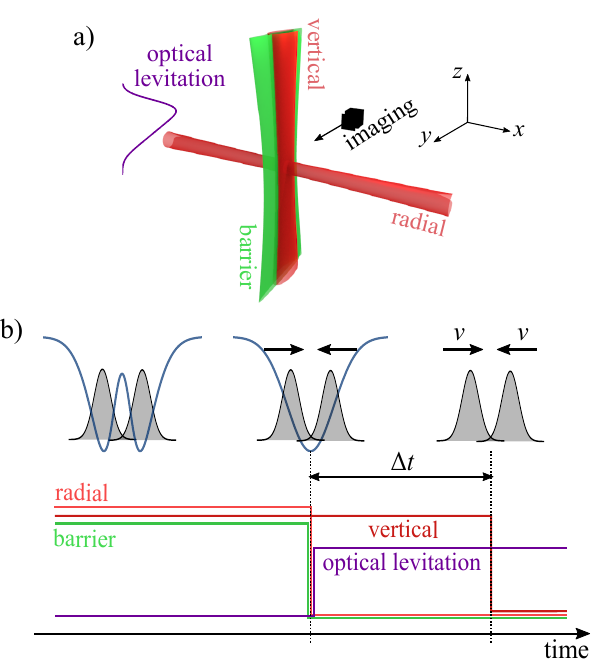}
\end{center}
\caption{Production of two colliding droplets. a) Schematic representation of the geometry of the optical potentials. b) Experimental sequence used to provide a controlled velocity to the droplets.}  
\label{fig1}
\end{figure}

The experimental setup is analogous to the one described in Ref. \cite{semeghini_2018}, with a modified trap geometry adapted to the creation of two colliding droplets. We first create two separate Bose-Einstein condensates (BECs) of $^{39}$K atoms in the $\ket{1,-1}$ hyperfine state (state 2) in a double-well potential, created by the superposition of a crossed dipole trap with a repulsive thin barrier that splits the BEC along the $x$ direction (Fig.\ref{fig1}a). By applying a radio-frequency (RF) pulse of 10 $\mu$s, we transfer $\sim$50$\%$ of the atoms in the $\ket{1,0}$ state (state 1), so as to create the desired attractive mixture. As already shown in Ref. \cite{cabrera_2018} and Ref. \cite{semeghini_2018}, in a specific range of magnetic fields $B$, the Feshbach resonances of these two hyperfine states are such that the intraspecies scattering lengths $a_{11}$ and $a_{22}$ are positive, while the interspecies $a_{12}$ is negative. When the effective scattering length $\delta a=a_{12}+\sqrt{a_{11}a_{22}}$ becomes negative, for $B<B_c=56.85$ G, the attractive mean-field energy would lead to a collapse of the BEC, while the repulsive energy provided by quantum fluctuations, the so-called Lee-Huang-Yang energy \cite{lhy_1957}, stabilizes the mixture and leads to the formation of self-bound atomic clouds. At the end of the RF pulse we thus have two quantum droplets separated by the repulsive barrier. In order to provide them a controlled and tunable velocity, we apply the following strategy (sequence sketched in Fig.\ref{fig1}b). We first switch off the radial dipole trap and the repulsive barrier, leaving the atoms in the vertical dipole trap plus the optical levitating potential introduced in \cite{semeghini_2018}. Due to the harmonic confinement provided by the vertical beam along the $x$ direction ($\omega_x=2\pi \times 93(5)$ Hz), the droplets move towards the center of the trap, acquiring an increasing velocity. After a time interval $\Delta t$ we switch off the vertical beam and the two clouds keep moving towards each other along the $x$ direction, at a constant velocity $v$. The value of $v$ is tuned by changing $\Delta t$: for $\Delta t< \pi/(2 \omega_x)$, increasing $\Delta t$ corresponds to a larger $v$. Via absorption imaging along the $y$ direction, we record the density profiles after a variable waiting time and observe the collisional dynamics.  

\begin{figure}[h] 
\begin{center}
\includegraphics{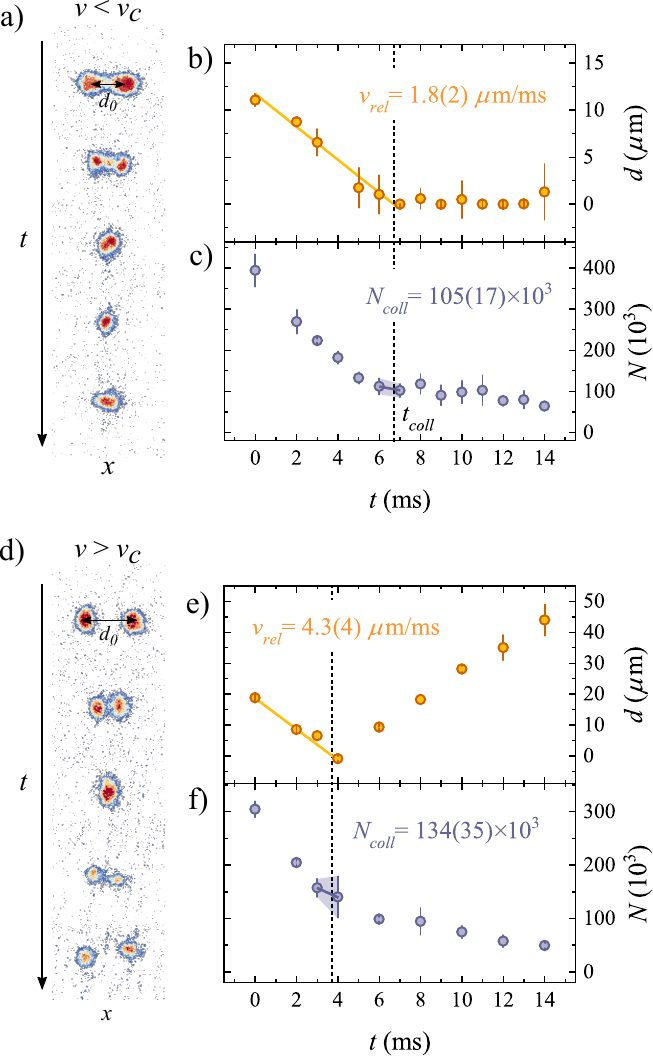}
\end{center}
\caption{Examples of two collision measurements resulting in merging (a) and separation (d) of the droplets. In b) and e) we report the corresponding evolution of the distance $d$ between the droplets and in c) and f) of the total atom number $N$. A linear fit of $d(t)$ before collision is used to measure $v_{rel}$ and $t_{coll}$. $N_{coll}$ is then deduced from a linear interpolation between the two data points adjacent to $t_{coll}$, as shown in c) and f). The data correspond to the average over four experiment repetitions. The error bars represent the statistical uncertainty and correspond to one standard deviation.}  
\label{fig2}
\end{figure}

We distinguish two different outcomes of the collision, as reported in Fig.\ref{fig2}a,d. When the droplets collide with velocities smaller than a critical value $v_c$, they merge in a single droplet, while for $v>v_c$ they separate after collision and keep moving in opposite directions. To measure the velocity of the collision, we fit the density profiles as a function of time $t$, using either a two-dimensional (2D) double-gaussian or a single-gaussian function, depending on whether there are two or one visible density peaks. In the former case we measure the distance between the droplets as $d=|x_L-x_R|$, where $x_L$ and $x_R$ are the centers of the left and right gaussians along the $x$ direction. In the single gaussian case, we define $d=0$. In Fig.\ref{fig2}b,e we plot $d(t)$ for the two experimental sequences reported in Fig.\ref{fig2}a,d: in the merging case $d$ stabilizes to zero after collision, while it linearly increases with time when the droplets separate. Performing a linear fit of $d(t)$ in the early stages of the collision, when the droplets are approaching each other, we can measure the relative velocity $v_{rel}$ and thus the velocity of each droplet $v=v_{rel}/2$ (Fig.\ref{fig2}b,e). Another relevant parameter to characterize the observed collisional dynamics is the total atom number at collision $N_{coll}$. We estimate the collision time as $t_{coll}=d_0/v_{rel}$, where $d_0=d(t=0)$, and then we determine $N_{coll}=N(t_{coll})$ by performing a linear interpolation between the two closest data points, as in Fig.\ref{fig2}c,f.

\begin{figure}[t]
\begin{center}
\includegraphics{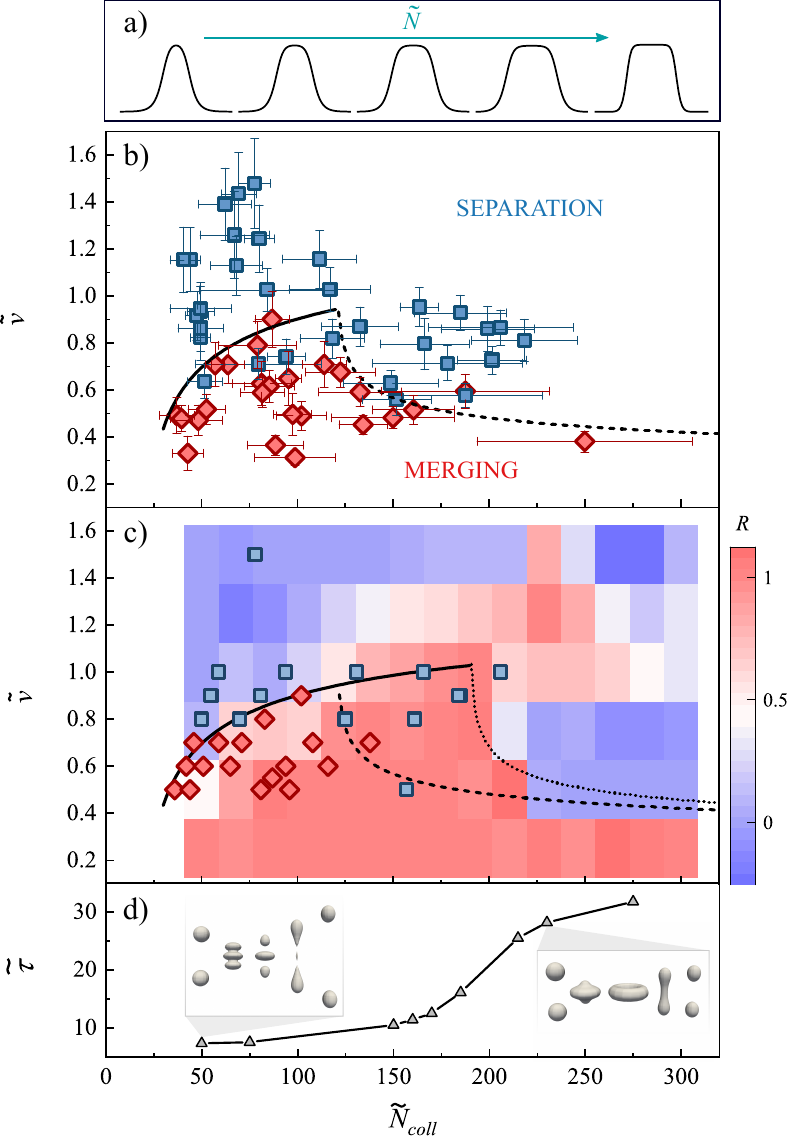}
\end{center}
\caption{Outcomes of the collision measurements (b) and simulations (c) as a function of $\tilde v$ and $\tilde N_{coll}$. In a) we draw the droplet wavefunction for increasing values of $\tilde N$, which shows the crossover from compressible to incompressible droplets. In c) the results of simulations in the ideal case without losses are represented as a color plot of the ratio $R$ introduced in the text, while the data points represent the results of the simulations with 3BL, distinguishing between merging (red diamonds) and separation (blue squares). The solid lines in b) and c) represent the expected trend $\tilde v_c \propto \sqrt{2|\tilde E_{drop}|/\tilde N}$ at small $\tilde N$, while the dotted in line in c) is $\tilde v_c\propto (\tilde N-\tilde N_0)^{-1/6}$, which is the predicted scaling at large $\tilde N$. The dashed lines in b) and c) correspond to the same $\tilde N^{-1/6}$ scaling, but in this case they are simply used as a guide to the eye. In d) we report the timescale of the collision $\tilde{\tau}$ as a function of $\tilde N_{coll}$ and, in the insets, two examples of the collisional dynamics observed in the simulations without 3BL, in the two opposite cases of small and large $\tilde N$. A detailed description of the estimation of the error bars in b) is reported in \cite{SI}.
}
\label{fig3}
\end{figure}

We take several datasets as a function of $v$ and $N_{coll}$. Note that, due to strong three-body losses in the system (as visible in Fig.\ref{fig2}c,f and already described in \cite{semeghini_2018}), we can tune $N_{coll}$ by changing the initial distance between the droplets and thus the collision time $t_{coll}$. In the sequences reported in Fig.\ref{fig2}, for example, in order to have a similar $N_{coll}$ in spite of the different $v$, $d_0$ was increased proportionally to $v$. In order to explore a broader range of atom numbers, we can also tune an additional parameter, i.e. the magnetic field $B$. As shown in \cite{petrov_2015}, the proper variable for the description of quantum droplets is indeed the rescaled atom number 
\begin{equation}
\tilde{N}=\frac{N }{(1+1/\alpha) n^{(0)}_1\xi^3}, \hspace{4mm} \textnormal{with} \hspace{4mm} \xi^2=\frac{3\hbar^2}{2m}\frac{1+\alpha}{|\delta g|n_1^{(0)}}
\label{eq:Ntilde}
\end{equation}
where $\alpha=\sqrt{a_{22}/a_{11}}$, $\delta g=4 \pi \hbar^2 \delta a/m$ and $n_1^{(0)}$ is the equilibrium density for the atoms in state 1, whose definition is reported in \cite{SI}.
$\tilde N$ defines the shape of the droplet wavefunction, thus distinguishing between a compressible regime, where there is no distinction between the bulk and the surface, and an incompressible regime, where the wavefunction displays a clear flat-top at the center, indicating the existence of a bulk with a fixed saturation density (Fig.\ref{fig3}a). $\tilde N$ depends on $B$ via the scattering lengths $a_{ij}$, so that we can control $\tilde N_{coll}$ by tuning $B$ in a range between 56.23 G and 56.44 G. 
Using the same rescaled units introduced in Eq.(\ref{eq:Ntilde}), the velocity of each droplet becomes $\tilde{v}=v m \xi/\hbar$.
In Fig.\ref{fig3}b we report the results of the collision measurements as a function of $\tilde N_{coll}$ and $\tilde v$, distinguishing between the two different outcomes: merging (red diamonds) and separation (blue squares). 
We observe a non-monotonic behavior of the critical velocity $\tilde v_c$, setting the threshold between the two regimes: for small droplets, $\tilde v_c$ increases with $\tilde{N}_{coll}$, while for larger droplets the trend is inverted.

In order to get a deeper insight in the physics of the collision and understand the observed behavior, we simulate numerically the dynamics of the collision by means of a modified Gross-Pitaevskii equation (GPE), equivalent to a time-dependent density functional theory within local density approximation \cite{ancilotto_2018}.
Similar approaches have been recently used to study both dipolar and mixture droplets \cite{chomaz_2016,baillie_2016,semeghini_2018}. Their validity for the ground state was also tested against Monte-Carlo simulations \cite{macia_2016,cikojevic_2018}. In our case, we assume the density ratio $n_1/n_2$ to be frozen, which is justified by the experimental results reported in \cite{semeghini_2018}, where a fast stabilization of $n_1/n_2$ to its equilibrium value was observed. 
We can then write an equation for a single macroscopic wavefunction~\cite{petrov_2015}. 
Using the rescaled units $\tilde{\bf{r}}=\bf{r}/\xi$ and $\tilde{t}=\hbar t/m\xi^2$, the modified GPE reads simply
\begin{equation}
 i\frac{\partial}{\partial \tilde t}\psi=-\frac{1}{2}\Delta_{\tilde{\bf{r}}} \psi-3|\psi|^2\psi+\frac{5}{2}|\psi|^3\psi .
 \label{mGPE}
\end{equation}
The initial state is represented by two quantum droplets with $\tilde N/2$ atoms each, separated by a certain distance to ensure that no overlap exists between them. We provide them a certain velocity $\tilde v$, so that they move towards each other, and then we observe the result of the collision. 
In order to distinguish between merging and separation, we measure, at large times after the collision, the ratio $R=n_{cm}/(n_{cm}+n_{out})$, where $n_{cm}$ is the density at the center of mass and $n_{out}$ is the peak density of the outgoing clouds. The results are reported in Fig.\ref{fig3}c. We observe a qualitative agreement with the experimental outcomes of Fig.\ref{fig3}b, but a quantitative difference in the position of the maximum of $\tilde v_c(\tilde N)$, which is here shifted to $\tilde N\sim 200$. To understand the origin of this discrepancy, we perform a new set of simulations including the effect of three-body losses (3BL). They are inserted in Eq.~(\ref{mGPE}) as an imaginary term $-i\frac{\tilde{K}}{2}|\psi|^4\psi$, where the parameter $\tilde{K}=0.53$ is determined by fitting the experimental atom number decay. In Fig.\ref{fig3}c we compare the new results (red diamonds and blue squares) with those obtained for the ideal case of no losses. While at small $\tilde N$ the position of $\tilde v_c$ is basically unaffected by losses, at larger $\tilde N$, 3BL shift the maximum of $\tilde v_c$ to $\tilde N\sim 120$, in good agreement with the experimental results. 

We can qualitatively understand the two opposite trends of $\tilde v_c (\tilde N)$ at small and large $\tilde N$ and the effect of losses, by drawing a simple argument. The possibility of forming a single droplet during the collision is related to the capability of the resulting merged droplet to absorb the excess kinetic energy $\tilde E_{kin}\propto \tilde{v}^2 \tilde{N}$. In the liquid regime at large $\tilde{N}$, we can decompose the energy of the droplet using the so-called liquid-drop model \cite{boronat_2009}: $\tilde{E}_{drop}(\tilde{N})=E_B\tilde{N}+E_S\tilde{N}^{2/3}+E_C\tilde{N}^{1/3}$, where the three terms represent the bulk, surface and curvature energies respectively. The first and the last term can be neglected, since the bulk energy scales linearly with $\tilde{N}$, and is thus conserved during the collision, and the curvature energy is negligible for large $\tilde N$. The only relevant energy scale is thus provided by the surface, which can also host discrete excitations to absorb the collision kinetic energy. Analogously to the Weber number criterion for classical liquid droplets \cite{qian_law_1997}, we can thus conclude that the condition for merging should be given by $\tilde E_{kin}\lesssim E_S \tilde N^{2/3}$, which means $\tilde v_{c} \propto \tilde N^{-1/6}$. In the opposite case of small $\tilde N$, there is no distinction between the bulk and the surface and one would expect the whole binding energy of the droplet to be the relevant energy scale. By imposing $\tilde E_{kin}\sim \tilde E_{drop}$ one gets a critical velocity $\tilde v_{c} \sim \sqrt{2|\tilde E_{drop}|/\tilde N}$.
In Fig.\ref{fig3}c we compare these expected trends with the numerical simulations performed in the absence of 3BL and we find that these simple energetic considerations reproduce pretty well the observed behavior.  
The crossover from compressible to incompressible droplets, governed by the two different energy scales, is highlighted by the timescales of the collision. In the simulations without 3BL, we consider collisions with $\tilde v$ slightly above $\tilde v_c$ and we estimate the time interval $\tilde{\tau}$ between $\tilde t_{coll}$ and the time when the distance between the two density peaks along $x$ becomes larger than the radial size of the droplets, for different $\tilde N$ (Fig.\ref{fig3}d). We observe a clear slowing down of the collisional dynamics as the droplet enters the liquid regime, where the surface tension dominates. In that limit, our simulations show a behavior closely analogous to the reflexive separation known in the context of classical drops (see e.g. \citep{ashgriz_poo_1990}): after colliding, the two droplets form a single excited cloud for a certain time interval, but they eventually separate when the surface tension is not sufficient to compensate for the kinetic energy of the internal flow (right inset of Fig.\ref{fig3}d). In the compressible case, instead, the separation occurs on a much shorter timescale, since the two droplets basically pass through each other (left inset of Fig.\ref{fig3}d). The reason for the different importance of 3BL in the two regimes lies exactly in the different timescale of the collision. The longer $\tilde{\tau}$ in the incompressible regime implies that, in the presence of 3BL, the atom number decreases significantly during the relevant time interval, so that the final surface tension is reduced. This corresponds to a smaller $\tilde v_c$ and thus to a shift in the position of the backbending, as highlighted by the dashed lines in Fig.\ref{fig3}b,c, used in this case as a simple guide to the eye, since a proper scaling in the presence of 3BL is harder to deduce. 

As a final remark, it is worth mentioning that the experimental procedure used to prepare the two initial droplets is such that there is no definite relative phase between them. In order to reproduce the incoherent preparation of the experiment, in the numerical simulations we set the initial phase difference to zero, which corresponds to minimizing the effects of phase gradients during the collision, thus recovering the proper hydrodynamic equations. Studying the effect of a finite relative phase is an interesting perspective of this work and it is the subject of ongoing investigations. 

In conclusion we have shown that binary collisions are a good probe of the dynamical properties of 3D mixture droplets. By comparing our experimental results with the outcomes of numerical simulations, we found evidence of a crossover from compressible to incompressible quantum droplets driven by $\tilde N$. This is highlighted by the different trend of $\tilde v_c(\tilde N)$, which is well justified by simple energetic considerations.       
In the future it would be interesting to study the coalescence dynamics of two droplets colliding at very small velocities, which, in analogy to previous studies on helium clusters \cite{vicente_2000, maris_2003, ishiguro_2004}, could be a probe of their superfluid properties. One could also investigate the formation of vortices during the collision and characterize the collective excitations of the merged droplets. 
\\

\acknowledgments{We acknowledge insightful discussions with F. Ancilotto, F. Dalfovo, S. Stringari and our colleagues of the Quantum Gases group at LENS. A.G. and A.R. acknowledge funding by the Provincia Autonoma di Trento and by the FIS$\hbar$ project of the Istituto Nazionale di Fisica Nucleare. This work was supported by EC-H2020 Grant QUIC No. 641122 and by the project TAIOL of QuantERA ERA-NET Cofund in Quantum Technologies (Grant Agreement N. 731473) implemented within the European Union's Horizon 2020 Programme.}

\section{SUPPLEMENTARY INFORMATION}

\subsection{Rescaled units and droplet equilibrium densities}
The rescaled units $\tilde{\bf{r}}$, $\tilde t$ and $\tilde N$ used in the paper were first introduced in \cite{petrov_2015}. 
They depend on the interatomic scattering legths $a_{11}$, $a_{22}$ and $a_{12}$ and on the equilibrium density of the atoms in state 1 in the droplet $n_1^{(0)}$. This was also derived in \cite{petrov_2015} as: 
\begin{equation}
n_1^{(0)}=\frac{25\pi}{1064}\frac{(a_{12}+\sqrt{a_{11}a_{22}})^{2}}{a_{11}a_{22}\sqrt{a_{11}}(\sqrt{a_{11}}+\sqrt{a_{22}})^5}.
\end{equation}
The density corresponding to the other hyperfine state is then determined at equilibrium by the minimization of the hard modes, which locks the ratio $n_1/n_2$ to $\alpha=\sqrt{a_{22}/a_{11}}$. 

\subsection{Double-well potential}
The double-well potential used to create two separate droplets is generated by the superposition of an infrared elliptical laser beam at 1064 nm and a thin green barrier at 532 nm. The infrared beam has an aspect ratio $w_x/w_y\sim 0.4$, while the green barrier is tightly focused along $x$, with $w_x=14(5)$ $\mu$m, and large along $y$ ($w_y=1.2(1)$ mm) to ensure the homogeneity of the repulsive potential along that direction. 
The intensity of the infrared beam is always set to the same value, so that $\omega_x^{IR}=2\pi \times 93(5)$ Hz. The height of the green barrier is instead changed in the different measurements, in order to tune the initial distance between the droplets $d_0$. The corresponding trapping frequencies in the single wells vary between 100 and 150 Hz.

\begin{figure}[h] 
\begin{center}
\includegraphics[width=\columnwidth]{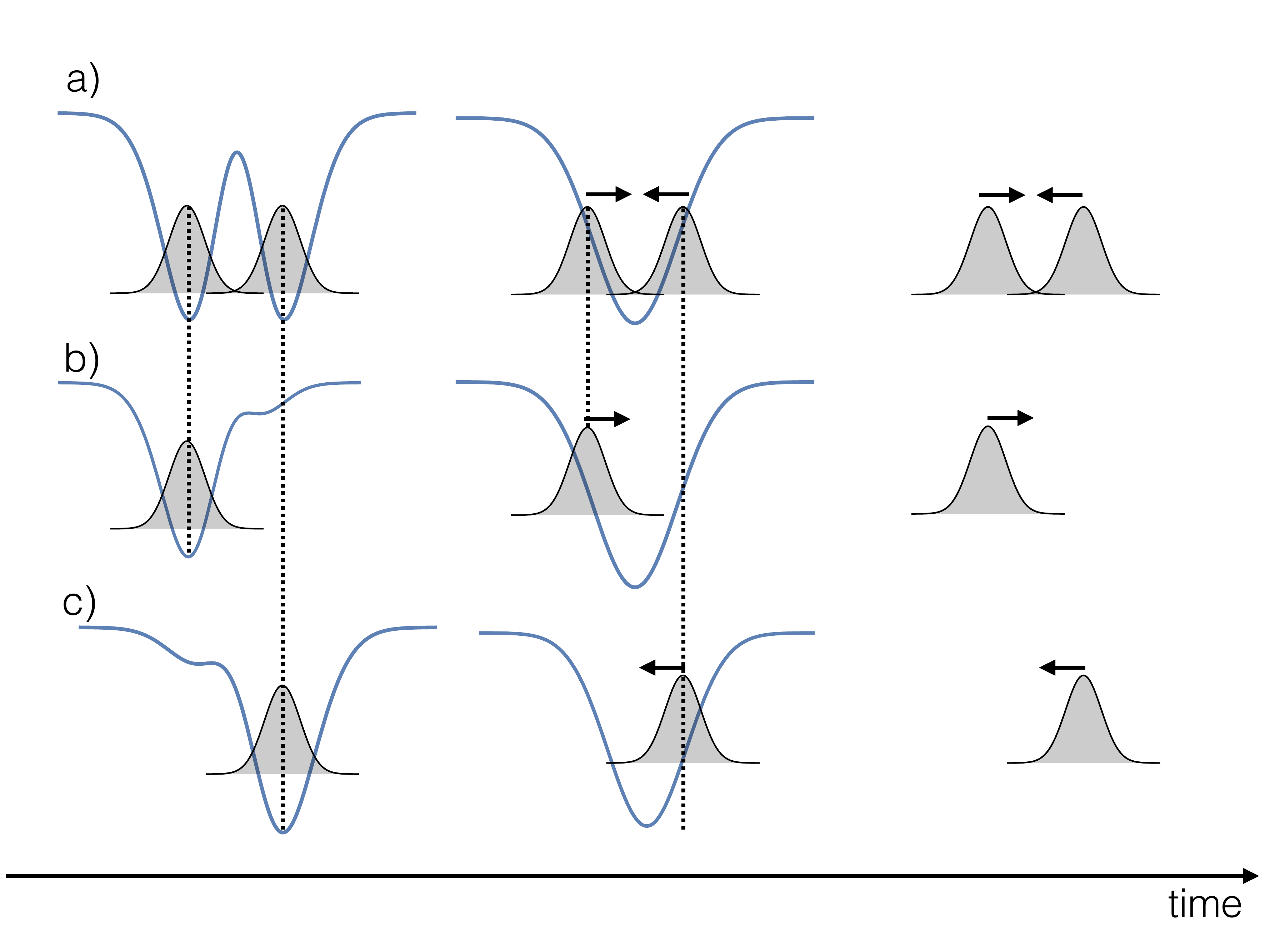}
\end{center}
\caption{a) Sketch of the experimental sequence used to provide a controlled relative velocity to the droplets. In b) and c) we load the BEC in an imbalanced double-well potential so that we create a single droplet at the same position of either the left or right droplet in the corresponding collision measurement (left column). Following the same sequence as in a) we can then provide the same velocity as in the collision measurement to each isolated droplet and obtain and independent measurement of $v_{rel}$.}  
\label{figS1}
\end{figure}

\subsection{Measurement of the droplet velocity}
The relative velocity $v_{rel}$ during the collision is measured, as explained in the main paper, by performing a linear fit of $d(t)$ when the droplets are approaching each other. In order to have an independent measurement of $v_{rel}$, where the trajectory followed by each droplet is not influenced by the presence of the other, we use the following strategy (sketched in Fig.\ref{figS1}). The basic idea is to repeat the sequence of Fig.1b, but having a single droplet at a time. In order to load the BEC in a single well, we increase the barrier height and we move it away from the center of the infrared beam, so that the resulting double-well potential is imbalanced and all the atoms occupy a single well, placed at the same position of the corresponding left or right well of the collision measurement (left column of Fig.\ref{figS1}). We then follow the same sequence as in Fig.1b and measure $x_L(t)$ and $x_R(t)$ independently. We evaluate the distance $d(t) = x_R(t) - x_L(t)$ and we deduce $v_{rel}$ from a linear fit to the data. We compare the results of this independent measurement with those obtained from the collision measurements and we observe that there is no systematic deviation between the two. We then conclude that measuring the droplets velocity from the initial dynamics of the collision, as we describe in the main paper, is a reliable strategy, since, before colliding, the trajectory of each droplet is not significantly affected by the presence of the other.

\subsection{Error bars in Fig.3}

The vertical error bars in Fig.3b of the paper are estimated considering two main sources of errors in the estimation of $\tilde v$, i.e. the uncertainty coming from the linear fit of $d(t)$ in Fig.2b,e and the uncertainty on the magnetic field $B$, which enters the definition of $\tilde v$. 
The horizontal error bars take into account the uncertainty coming from the linear interpolation used to determine $\tilde N_{coll}$ (Fig.2c,f), the uncertainty on the magnetic field $B$ which enters the definition of $\tilde N$ (Eq.1) and the uncertainty on the calibration of the atom number $N$.
All the error bars correspond to one standard deviation.

\end{document}